# NEUTRON CAPTURE NUCLEOSYNTHESIS

**Miklós Kiss**

Berze High School, Gyöngyös, Hungary, mikloskiss2630@gmail.com

**ABSTRACT**

*Heavy elements (beyond iron) are formed in neutron capture nucleosynthesis processes. A simple unified model is proposed to investigate the neutron capture nucleosynthesis in arbitrary neutron density environment. Neutron density required to reproduce the measured abundance of nuclei assuming equilibrium processes is investigated as well. Medium neutron density was found to play a particularly important role in neutron capture nucleosynthesis. Using these findings most of the nuclei can be formed in a medium neutron capture density environment e.g. in certain AGB stars. Besides these observations the proposed model suits educational purposes as well.*

**INTRODUCTION**

Nearly sixty years after BBFH [1], it is possible and necessary to review and rethink our knowledge about the neutron capture nucleosynthesis. The result of the formation of the nuclei is shown in the various abundances. It is important to mention that the unstable nuclei decayed into stable nuclei and we are only able to observe the abundance of the remaining stable nuclei. "The success of any theory of nucleosynthesis has to be measured by comparison with the abundance patterns observed in nature." – say Käppeler, Beer and Wisshak [2], that is, we need to create such model that gives back the observed abundances.

Since the formation of nuclei takes place in a variety of conditions, the experienced abundance is a result of several processes. Therefore, more models are necessary for different conditions. According to the conditions of the models the nuclei are classified into categories as s-nuclei, r-nuclei etc.

It seems that the reverse approach is also useful: the observed abundances preserve the conditions of the formation of the nuclei. Instead of investigating whether the theoretical model fits the observed abundance, we look for the circumstances when the observed abundance is known. To do this, we need suitable data: the half-life of unstable nuclei and the neutron capture cross section of nuclei. These data are not always constant. For certain nuclei the half-life depends on the temperature [2], [3] and [4]. Fortunately, the reaction rate per particle pair $<\sigma v>$ is constant between 10 and 100 keV because of the energy dependence of $\sigma$ [2], [3]. So we can use the $\sigma$ values at 30 keV [5]. The possible resonances only improve the capture capabilities.

**OVERVIEW**

Many nuclei are known (Fig.1.). Processes that are used to describe the formation of nuclei of elements heavier than iron were defined by Burbidge, Burbidge, Fowler and Hoyle (B2FH) in 1957 [1]. In sixty years the model was refined such that element abundances in the Solar System are reproduced with less than one percent error. Such accurate quantitative description leads to the general and unquestioned acceptance of two main processes that describe neutron capture nucleosynthesis in the literature: the s-process (slow process) in low neutron density environments such as helium- and carbon-oxygen-burning asymptotic giant branch stars and the r-process (rapid process) in high neutron density environments, typically supernova explosions. This separation relies on the fact that an unstable nucleus can either decay or first capture another neutron. Assuming individual nuclei and s-process a neutron capture is expected every ten years. Nuclei having a half-life less than a year almost certainly decay. Formation of elements occurs along a path (s-path) in the nuclear valley of stability, from the light towards the heavy elements. Only nuclei having a lifetime longer than ten years must be considered if the strictly classical approach is followed. So the question arises: Does the neutron take part in the neutron capture process during its full fifteen-minute lifetime at all?

Obviously, the classical approach needs refinement if the neutron capture time and decay time are of the same order. The notion of branching was introduced for such cases in the s-process [2] and even the latest literature relies on this view [5], [6]. The r- and s-nuclei are important observation evidences of the two processes. According to the classical model, s-nuclei can only form in s-processes and r-nuclei can only form in r-processes.

In the classical model nucleosynthesis by the s-process occurs in a path along the valley of beta stability and it is generally accepted that the s-process nucleosynthesis terminates at bismuth by the fast alpha decay of polonium [7]. Only with the r-process is it possible to go further from the valley of beta stability to the neutron rich region of nuclei, towards the neutron drip line. And only by the r-process is it possible bypass the trap of polonium.

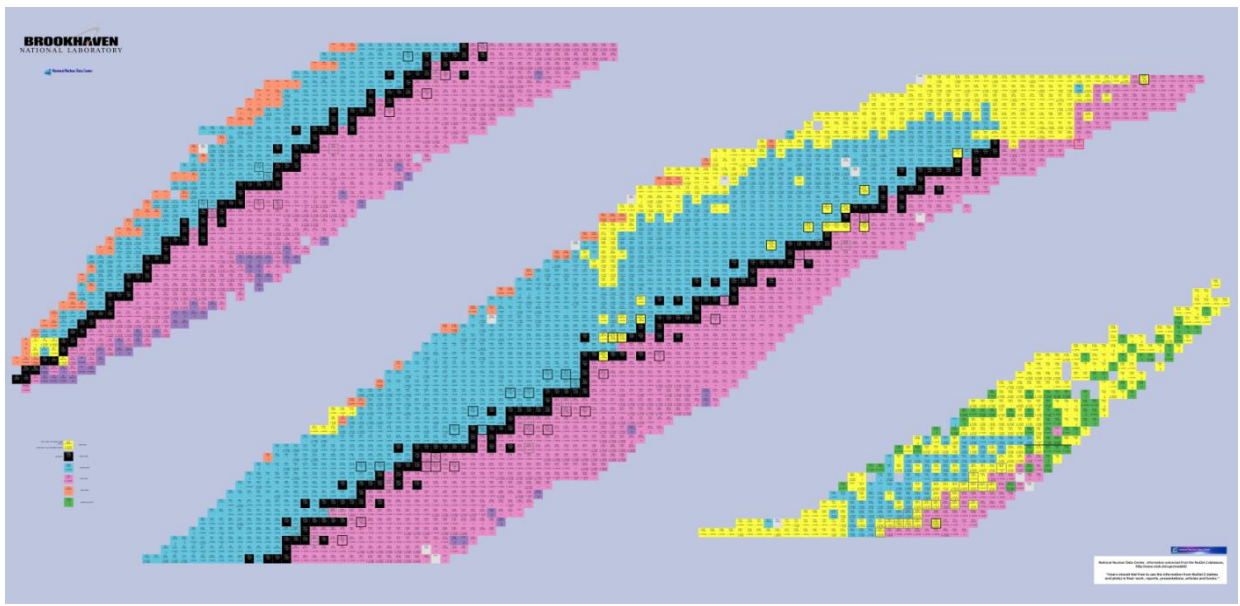

Fig.1. The Chart of Nuclei [8]

**THE NUCLEOSYNTHESIS BASED ON OUR MODEL**

A computer simulation program was created in order to make a graphical representation of nucleosynthesis by neutron capture [9-12]. The illustrative presentation made the formation and abundance of nuclei visible. Previously unknown details of nucleosynthesis became observable. In the differential equations that provide the basis of our model, the nuclei are distinguished by both atomic and neutron number. The possible values of the neutron density were not limited. The formation of elements is determined by the half-life, the neutron capture cross section, the neutron density and the amount of their parent's nuclei (overstep threshold). The model allows to investigate the role of the individual parameters in the formation of nuclei as well.

Our model requires specific nuclear data for each individual nucleus. Some of these (such as half-life, decay mode and branching ratio) can be found in the literature, but neutron capture cross sections were missing in some cases, mostly for those nuclei that are not used in the classical s- or r-processes. Due to the lack of these data we have investigated both measured and calculated neutron capture cross sections at 30 keV from different sources. Studying published values of Maxwellian averaged neutron capture cross sections we found that those obeyed simple phenomenological rules as a function of proton and neutron number. We found some simple rules for the location of the highest capture cross section on the Z-N plane (Ridge of neutron capture cross sections) and also its maximum value. We used these rules to make predictions for cross sections of neutron capture on nuclei with proton number above 83, where very few MACS data is available and needed for our model [13].

Exploiting the capabilities of our model we have investigated the formation of the elements using various assumptions. We have established that that nucleosynthesis of heavy elements occurs along a wide band near the valley of the beta stability. Our model, named band-process, is based on simple physical assumptions [9-12]. The width of the band and the isotope with the maxima amount in the band depend on the neutron density. The initial amount of nuclei is also important and it depends on the mass and metallicity of the star.

If we use a long time for step and a small amount initial iron nuclei, then we exclude the most of the nuclei from formation. Using a short time interval (one second or shorter) and sufficient amount of initial iron, we can see the true nature of the processes, the band formation of nuclei during the nucleosynthesis by neutron capture. The neutron capture process at low neutron density ($n_n \sim 10^7 - 10^8 \, cm^{-3}$) is called s-process and the neutron capture process at high neutron density ($n_n \sim 10^{20} - 10^{25} \, cm^{-3}$) is called r-process. Based on our model we can see what these concepts mean during nucleosynthesis and how the bands evolve in the different cases.

We have found that the processes that occur at moderate, $n_n \sim 10^{10} - 10^{14} \, cm^{-3}$ neutron density (m-process) are very important. They typically take place during the TP state of AGB stars [14], [15].

The nucleosynthesis in stars with moderate mass and low neutron density occurs in a band along the valley of beta stability and is terminated at Z=83 by the alpha decay of the polonium. If the neutron density exceeds $n_n = 10^{12} \, cm^{-3}$, the evolution does not terminate at bismuth. In AGB stars during a thermal pulse the neutron density significantly exceeds this value so the valley may proceed to fermium. The simulated abundance approximates well the observations, only the abundance of lead is higher than the expected value. Our program indicates the production of r-nuclei at low neutron density and s-nuclei at high neutron density. The r-nuclei found in SiC meteorite grains demonstrate the possibility of their formation in slow or moderate processes [5].

The band of nucleosynthesis reaches the adjacent r-nuclei even at low neutron density, although the amount of these nuclei is negligible. At moderate neutron density the amount of these nuclei increases to the empirical abundance found in the Solar System. However, the experienced abundance can be reproduced with nuclei produced at processes with low, intermediate and high neutron density.

In the case of high neutron density, if the neutron density does not significantly exceed the $n_n = 10^{20} \, cm^{-3}$ value then some s-only nuclei are also produced with some exceptions ($^{176}_{72}Hf_{104}$ and $^{192}_{78}Pt_{114}$) [9].

We found that neutron capture process in the AGB TP phase at intermediate neutron density forms elements heavier than bismuth, the formation of nuclei can evolve even to fermium [9]. Although "sweep-out" obscures this situation, there is an empirical argument that confirms the predictions of our model. This argument is the anomaly of the isotopic abundance of tellurium, namely the two most abundant tellurium isotopes are r-nuclei with thirty-thirty percent abundance. This anomaly is unique; there are no other elements with such a strange isotopes anomaly. It seems that there are two arguments: 1. the formation band which includes these two nuclei, 2. fermium that forms in the AGB TP phase goes through spontaneous fission and results in unstable tin isotopes. The unstable tin isotopes decay into $^{128}_{52}Te_{76}$ and $^{130}_{52}Te_{78}$, and that is the other way part of these two abundant isotopes come into existence [9]. We can reproduce the distribution of tellurium isotopes with a linear combination of isotope distributions obtained with slow, moderate and rapid processes. According to our model the most important places of element formation are AGB stars. This is in agreement with the recent results in the literature [6], [15]. The observation of radio nuclei $^{26}Al$ and $^{60}Fe$ in the Milky Way and the discovery of the daughter isotopes in the pre-solar grains provide further confirmation of the conclusions of our model [5], [16].

**VERIFICATION**

Currently, verification of a model almost completely relies on one criterion: the abundances of elements in the Solar System [17] should be correctly reproduced. However, observed abundances include the aggregate effects of multiple processes that take place during nucleosynthesis. It seems reasonable to assume the existence of an intermediate neutron density nucleosynthesis to bridge the gap between the s- and r-processes. Anomalous isotopic ratios observed in early meteorites substantiate such assumptions [5]. Intermediate processes take place in AGB stars.

There are several ways to verify the model. Besides the ratio of r- and s-nuclei, mainly the accurate reproduction of the abundances measures the goodness of the model. The latter however, depends on many other parameters, because the elements form in stars of different states under different conditions. The abundance of isotopes is very important for verification as well. Discovery of radiogenic nuclei with long half-life in the Milky Way or elsewhere in the universe could be an important evidence, too.

An independent way of verification is the rate analysis [18]. Abundances of elements can be classified as elemental abundance, isotopic abundance, and abundance of nuclei. In our works the nuclei are identified by (Z,N), which allows reading out new information from the measured abundances. We are interested in the neutron density required to reproduce the measured abundance of nuclei assuming equilibrium processes. This is only possible when two stable nuclei are separated by an unstable nucleus. At these places we investigated the neutron density required for equilibrium nucleosynthesis both isotopically (Fig.2.) and isotonically (Fig.3.) at temperatures of AGB interpulse and thermal pulse phases. We obtained an estimate for equilibrium nucleosynthesis neutron density in most of the cases. Next we investigated the possibility of partial formation of nuclei. We analyzed the meaning of the branching factor. We found a mathematical definition for the unified interpretation of a branching point closed at isotonic case and open at isotopic case. We introduced a more expressive variant of branching ratio called partial formation rate. With these we were capable of determining the characteristic neutron density values. We found that all experienced isotope ratios can be obtained both at $10^8 K$ temperature and at $3 \cdot 10^8 K$ temperature and at intermediate neutron density ($\leq 2 \cdot 10^{12} cm^{-3}$).

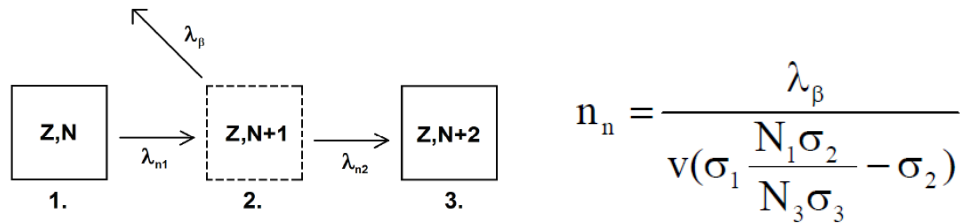

Fig.2. The isotopic case and its equilibrium neutron density

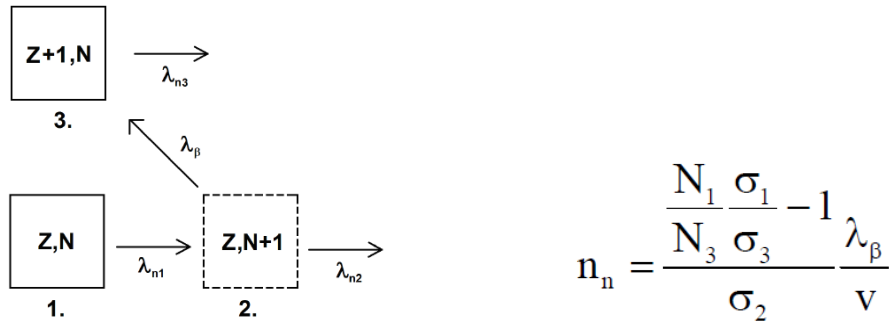

Fig.3. The isotonic case and its equilibrium neutron density

**EDUCATIONAL USAGE**

The origin of heavy elements beyond iron is an important and interesting problem; therefore, it is worth discussing it in the classroom. Recently these questions are presented in the final exam of high schools [19].

Traditionally we speak only about fusion and decays, but it also is an interesting question how the elements heavier than iron are formed. The primary purpose of our investigation was educational. With the model the processes can be followed and demonstrated easily and can be integrated into education.

Our model is capable of demonstrating the decay processes, the decay trees from arbitrary initial nucleus (i.e. isotope) as well. So we can follow the change of the amount of seed isotopes and the daughter isotopes as well.

**CONCLUSIONS**

The neutron capture formation of nuclei occurs in a band. There are no r-nuclei (in exclusive meaning). Most s-nuclei can form in r-process. The bypass of bismuth is possible at medium neutron density. All experienced isotope ratios can be obtained both at $10^8$ K temperature and at $3 \cdot 10^8$ K temperature at intermediate neutron density ($10^{12}$-$10^{14}$ cm$^{-3}$), so the m-process and the AGB stars are probably one of the main places of nucleosynthesis. It seems that the so-called r-nuclei can form in intermediate processes as well.

Our model is also capable of visualizing the processes of neutron capture nucleosynthesis and the decay processes as well.

**ACKNOWLEDGMENTS**

I am grateful to Zoltán Trócsányi who provided assistance in developing the theme during my PhD studies. I am grateful to Zsolt Fülöp and the Atomki as well, for his continuous support of my work.